\documentclass[10pt,notitlepage,aps,prc,longbibliography,nofootinbib,floatfix]{revtex4-1}
\usepackage{graphicx,bm,xcolor, bbold}
\usepackage[normalem]{ulem}
\usepackage{hyperref}
\hypersetup{colorlinks, linkcolor={blue},citecolor={blue},urlcolor={blue}}
\usepackage{graphicx}
\usepackage{amsmath,amssymb,amsfonts}
\usepackage{bm}
\usepackage{color}
\graphicspath{fig}
\usepackage{epstopdf}
\usepackage{multirow}
\usepackage{scalerel}

\newcommand{\new}[1]{{\color{red}{#1}}}

\newcommand{\remove}[1]{\ifmmode\mbox{\new{\sout{$#1$}}}\else\new{\sout{#1}}\fi}

\newcommand{\vect}[1]{\bm{\mathrm{#1}}}

\newcommand{\jv}{\vect{j}}
\newcommand{\kv}{\vect{k}}

\newcommand{\pv}{\vect{p}}

\newcommand{\vv}{\vect{v}}

\newcommand{\nv}{\vect{n}}

\begin{document}

\title{Emergence of the geometric contribution to the superfluid density in the inner crust of neutron stars}
\author{Giorgio Almirante}
\email{almirantecomma@live.it}
\affiliation{Universit\'e Paris-Saclay, CNRS/IN2P3, IJCLab, 91405 Orsay, France}

\begin{abstract}
The geometric contribution to the superfluid density has been found to be of great importance in the inner crust of neutron stars. In this work we clarify how this contribution arises in the context of a band theory for neutrons. Specifically, we derive the dependence of the superfluid density on the magnitude of the pairing gap when the system has many bands cutting the Fermi energy, as it is the case for the neutrons in the inner crust. Also, in the perturbation theory framework, we find that it is essential to account for the corrections to the (Bogoliubov) quasi-particle states in order to get the geometric contribution. Accounting only for the corrections to the (Hartree-Fock) single-particle states leads to the conventional contribution only.
\end{abstract}

\maketitle

\section{Introduction}
The inner crust of neutron stars is expected to be composed of clusters of neutrons and protons, surrounded by a gas of unbound neutrons and a background gas of relativistic electrons, ensuring charge neutrality and $\beta$-equilibrium \cite{Chamel08}. The clusters probably form a periodic lattice due to the interplay between the short-range nuclear force and the long-range Coulomb interaction, while the surrounding neutron gas has densities at which pure neutron matter is superfluid. The superfluid component of the crust could have observable consequences for the hydrodynamical and thermodynamical properties of the star \cite{Page12}. Also, there is the belief that it is involved in the mechanism that produces pulsar glitches \cite{Zhou22}. To compare these models with observations, it is necessary to know some microscopic features of the inner crust \cite{Antonelli22}, such as the neutron superfluid density.

The crucial point in computing the actual superfluid density is the evaluation of the so-called ``entrainment'', that is a non-dissipative force between the superfluid component and the nuclear lattice \cite{Prix02,Carter06A}. In order to assess this quantity, various band-structure calculations for the neutrons have been performed in the absence of superfluidity \cite{Chamel05,Chamel06,Chamel12A,Kashiwaba19,Sekizawa22}, thus extracting the conduction density of neutrons. These calculations drew a picture where the neutrons were strongly entrained by the lattice. Then, first attempts to include pairing confirmed this picture, showing a very weak dependence of the superfluid fraction on the magnitude of the pairing gap \cite{Carter05A,Chamel24}.
However, recently band-structure calculations with self-consistent pairing have been performed for the inner crust, accounting for both the "pasta" phases \cite{Yoshimura24,Almirante24,Almirante24A} and the crystalline one \cite{Almirante25A}, resulting in a superfluid fraction much higher than previously expected. Actually, a problem was recognized in the formula for the superfluid density derived in \cite{Carter05A,Chamel24}, namely the neglect of the so-called geometric contribution to the superfluid density, which turned out to be of great importance in the inner crust of neutron stars \cite{Almirante25}.

The geometric contribution to the superfluid density has been studied in condensed matter systems since few years, both in ultra-cold atoms \cite{Peotta15,Liang17,Lukin23} and multi-band BCS superconductors \cite{Iskin24,Jiang24,Iskin25}. Most of these studies have concentrated on flat-band systems, i.e. when the conduction density in the normal phase is negligible, but recently some efforts have been done to understand how the geometric contribution arises from the underlying band structure \cite{Kitamura22,Hu25}.

The inner crust of neutron stars shares with condensed matter systems its periodic nature, thus one can treat both within the band theory framework. However, apart for the difference in the underlying interaction, these systems are characterized by very different density regimes. This is reflected in the different number of energy bands that one needs to treat in order to actually describe the system of interest. On one hand, in condensed matter accounting for few bands around the Fermi energy is enough to get a description of the system \cite{Liang17,Tian23}. On the other hand, the neutrons in the inner crust present a highly non-trivial band structure \cite{Chamel12A,Almirante25A}, with many bands cutting the Fermi energy.

In order to assess what differences one can expect in the study of such different systems, in this work we start to clarify how the geometric contribution to the superfluid density emerges when having a highly non-trivial band structure. In particular, the expression for the superfluid density found in \cite{Almirante25} is analyzed, and the framework developed there is applied to compute responses to a stationary flow under different approximations. Analyzing the expression found in \cite{Almirante25} for small values of the pairing gap, we show that the superfluid density has a linear dependence on the magnitude of the gap. This opens the way for a comparison between different systems: if a condensed matter system in which the superfluid density has this behavior is found, one can start to explore the possibility to have an experimental platform to mimic the physics of the inner crust of neutron stars. Then, computing the response of the system to a stationary flow under different approximations, on one hand we find the presence of two contributions in the response also in the absence of superfluidity, and on the other hand we identify the approximation performed to obtain the expression for the superfluid density derived in \cite{Carter05A,Chamel24}.

In Section \ref{sec:formalism} the formalism developed in \cite{Almirante25} is briefly recalled. In Section \ref{sec:emergence} it is shown how the geometric contribution emerges when many bands are involved. In Section \ref{sec:singleparticle} the formalism is applied to compute responses under different approximations. In Section \ref{sec:conclusions} there are the conclusions.

\section{Formalism} \label{sec:formalism}
\begin{figure}
    \centering
    \includegraphics[width=10cm]{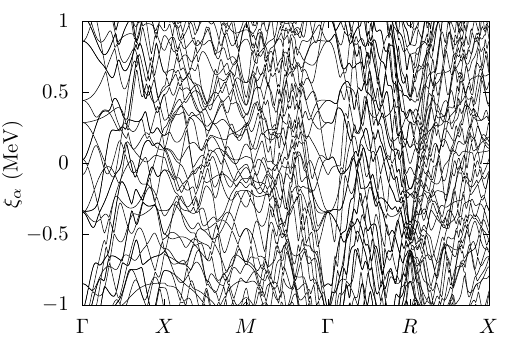}
    \caption{Single-particle band structure in the simple cubic lattice for a case example in the crystal phase of the inner crust of neutron stars (see \cite{Almirante25A} for details). $\xi_\alpha=\epsilon_\alpha-\mu$ are HF eigenvalues, the first $70$ bands around the Fermi energy are plotted. The horizontal axis is the Bloch momentum along the high symmetrical path of the simple cubic Brillouin zone \cite{Setyawan10}.}
    \label{fig:bands}
\end{figure}

We want to describe the crystalline phase of the inner crust of neutron stars, which is a fermionic 3D periodic system, within a mean-field approach. The basic quantity from which we start is thus the mean-field Hamiltonian $h$, which defines the Hartree-Fock (HF) eigenvalue problem
\begin{equation} \label{eq:HFeigen}
    h|\alpha\kv\rangle=\epsilon_{\alpha\kv}|\alpha\kv\rangle \,,
\end{equation}
where $\epsilon_{\alpha\kv}$ are the single-particle energies, and $|\alpha\kv\rangle$ are the single-particle states. The quantum number $\alpha$ labels the eigenvalues and eigenvectors of the HF eigenvalue problem. The $\kv$ dependence refers to the Bloch momentum defined in the first Brillouin zone (BZ). This is the essential ingredient to construct the band structure of the system: the Hamiltonian is diagonal in the Bloch momentum $\kv$, which encapsulates the periodic structure of the system \cite{Ashcroft}. In Figure \ref{fig:bands} the single-particle band structure is shown for a case example in the crystal phase of the inner crust of neutron stars, which is computed solving the HF eingenproblem in Eq.~(\ref{eq:HFeigen}) for different values of the Bloch momentum (see \cite{Almirante25A} for details).

Then the idea is to study the system in the presence of a stationary flow. In order to do this a perturbation $V'=-\vv\cdot\pv$ is added to the Hamiltonian, corresponding to a Galilean boost. Including the perturbation $V'$ means to study the system in the reference frame in which the mean-field has constant velocity $\vv$.

If in the unperturbed system there is no current, the average momentum density $\langle\jv\rangle$ of the perturbed system comes just from the perturbative correction to the density matrix. Writing the matrix elements of the perturbed density matrix in the single-particle basis as $\rho_{\alpha\beta\kv}+\rho'_{\alpha\beta\kv}$, this means that 
\begin{equation} \label{eq:current}
     \langle\jv\rangle = 2 \int_{\text{BZ}} \frac{d^3k}{(2\pi)^3} \sum_{\alpha\beta} \rho'_{\alpha\beta\kv}\pv_{\beta\alpha\kv} \,,
\end{equation}
where $\alpha$ and $\beta$ are single-particle quantum numbers and the factor $2$ accounts for the spin degeneracy. The momentum operator $\pv$ has been written as its matrix elements in the single-particle basis, and a useful way to rewrite them is the following
\begin{equation} \label{eq:momentum}
    \frac{\pv_{\alpha\beta\kv}}{m}=\frac{1}{m} \langle\alpha\kv|\pv|\beta\kv\rangle=
    \frac{\partial\xi_{\alpha\kv}}{\partial \kv} \delta_{\alpha\beta} +
    (\xi_{\alpha\kv}-\xi_{\beta\kv})\sum_{\nv}\frac{\partial\langle\alpha\kv|\nv\kv\rangle}{\partial \kv}\langle\nv\kv|\beta\kv\rangle\,,
\end{equation}
where $m$ is the bare nucleon mass, $\xi_\alpha=\epsilon_\alpha-\mu$ are the single-particle energies after subtracting the chemical potential $\mu$, and $|\nv\kv\rangle$ are momentum eigenstates. The momentum eigenstates are labeled by an integer number $\nv$ referring to vectors of the reciprocal lattice and the Block momentum $\kv$ defined in the BZ \cite{Ashcroft}. The relation for the matrix elements of the momentum operator Eq.~(\ref{eq:momentum}) is obtained performing the derivative with respect to the Bloch momentum $\kv$ of the matrix elements of the mean-field Hamiltonian $\langle\alpha\kv|h|\beta\kv\rangle$, and using Eq.~(\ref{eq:HFeigen}), accounting for the fact that the mean-field potential is independent on $\kv$.

At this point one can compute the correction to the density matrix and thus the average momentum density. Then, the average momentum density can be compared with its expression in the two-fluid model \cite{Andreev75}, that working in the rest frame of the superfluid and in the absence of momentum-dependent interactions, reads
\begin{equation} \label{eq:currenttwofluid}
    \langle\jv\rangle=(\langle\rho\rangle-\rho_S)m\vv \,,
\end{equation}
where $\langle\rho\rangle$ is the average density and $\rho_S$ is the superfluid density. In this way one can extract the superfluid density. The relation in Eq.~(\ref{eq:currenttwofluid}) is valid in the reference frame in which the superfluid is at rest and the mean-field has constant velocity $\vv$.

It has to be noticed that the above construction is independent on the way in which one compute the correction to the density matrix $\rho'$. The only assumptions are that the system can be described within a mean-field approach in which the single-particle labels are good quantum numbers, and that it is 3D periodic.

\subsection{Linear response in HFB}
In \cite{Almirante25} the correction to the density matrix is computed through perturbation theory applied to the Hartree-Fock-Bogoliubov (HFB) equations \footnote{The HFB equations are equivalent to the Bogoliubov-de Gennes (BdG) ones in condensed matter theory.}, starting from an unperturbed Bardeen-Cooper-Schrieffer (BCS) ground state (with BCS we refer to the case in which the pairing gap is constant). The BCS expressions for the quasi-particle energies $E_{\alpha\kv}$ and the coefficients $u_{\alpha\kv}$, $v_{\alpha\kv}$ read
\begin{equation} \label{eq:BCSEv}
    E_{\alpha\kv}=\sqrt{\xi^2_{\alpha\kv}+\Delta^2}
    ~~~,~~~
    v^2_{\alpha\kv}=\frac{1}{2}\Big(1-\frac{\xi_{\alpha\kv}}{E_{\alpha\kv}}\Big)
    ~~~,~~~
    u^2_{\alpha\kv}=\frac{1}{2}\Big(1+\frac{\xi_{\alpha\kv}}{E_{\alpha\kv}}\Big) \,,
\end{equation}
where $\Delta$ is the pairing gap.

Then, the perturbation discussed in Sec. \ref{sec:formalism} is applied, computing the first order perturbative correction to the HFB quasi-particle states (see \cite{Almirante25} for details), and the superfluid density results to be ($i,j=x,y,z$)
\begin{equation} \label{eq:supdens}
    \rho_S^{ij} = \int_{\text{BZ}}\frac{d^3k}{(2\pi)^3} \Bigg[m \sum_{\alpha} \frac{\Delta^2}{E_{\alpha\kv}^3} \frac{\partial\xi_{\alpha\kv}}{\partial k^i}\frac{\partial\xi_{\alpha\kv}}{\partial k^j}
    +\frac{2}{m}\sum_{\alpha\neq\beta} \frac{\Delta^2}{E_{\alpha\kv} E_{\beta\kv} (E_{\alpha\kv} + E_{\beta\kv})} 
    p^i_{\alpha\beta\kv} p^j_{\beta\alpha\kv}
    \Bigg] \,,
\end{equation}
where one can recognize a first term which sums the single bands, know as the conventional contribution, and a second term coming from the mixing of different bands, the so-called geometric contribution, namely
\begin{align} \label{eq:supdens_conv}
    \rho_{S~\text{conv}}^{ij} &= m \int_{\text{BZ}}\frac{d^3k}{(2\pi)^3} \sum_{\alpha} \frac{\Delta^2}{E_{\alpha\kv}^3} \frac{\partial\xi_{\alpha\kv}}{\partial k^i}\frac{\partial\xi_{\alpha\kv}}{\partial k^j} \,, \\
    \label{eq:supdens_geom}
    \rho_{S~\text{geom}}^{ij} &= \frac{2}{m} \int_{\text{BZ}}\frac{d^3k}{(2\pi)^3} \sum_{\alpha\neq\beta}
    \frac{\Delta^2}{E_{\alpha\kv} E_{\beta\kv} (E_{\alpha\kv} + E_{\beta\kv})} 
    p^i_{\alpha\beta\kv} p^j_{\beta\alpha\kv} \,.
\end{align}
%

\section{Emergence of the geometric contribution} \label{sec:emergence}
To get some insight in how the geometric contribution emerges when accounting for a finite pairing gap, let us show some limiting cases for the expression in Eq.~(\ref{eq:supdens_geom}). In particular, the focus will be on small pairing gaps, since the BCS approximation is expected to be valid in the weak-coupling limit, and because we want to understand how gap effects on the superfluid density emerge in relation to the band structure. In general, the band structure can be highly non-trivial, in the sense than many single-particle energies can be close to or cut through the Fermi energy, and thus small gaps do not mean that the single-particle bands cannot have energies smaller than the gap.

For simplicity in the following we will not write explicitly the dependencies on the Bloch momentum $\kv$, but it is important to notice that all the quantities will depend on the same $\kv$, apart from the pairing gap which is constant. 

In order to study the geometric contribution, one can rewrite the momentum operators in Eq.~(\ref{eq:supdens_geom}) using Eq.~(\ref{eq:momentum}), namely
\begin{equation} \label{eq:geometric}
    p^i_{\alpha\beta} p^j_{\beta\alpha} =
    (\xi_\alpha-\xi_\beta)^2\frac{\partial\langle\alpha|}{\partial k^i}|\beta\rangle\langle\beta|\frac{\partial|\alpha\rangle}{\partial k^j} =
    (\xi_\alpha-\xi_\beta)^2~g_{\alpha\beta}^{ij} \,,
\end{equation}
for $\alpha\neq\beta$, where we defined the tensor
\begin{equation}
	g_{\alpha\beta}^{ij}=\frac{\partial\langle\alpha|}{\partial k^i}|\beta\rangle\langle\beta|\frac{\partial|\alpha\rangle}{\partial k^j} \,.
\end{equation}
From this one can see why the term in Eq.~(\ref{eq:supdens_geom}) is called geometric contribution: the tensor $g_{\alpha\beta}^{ij}$ is related to the geometry of the manifold of the single-particle states \cite{Provost80}.

Now, inserting Eq.~(\ref{eq:geometric}) in Eq.~(\ref{eq:supdens_geom}), and multiplying both numerator and denominator by $(E_\alpha-E_\beta)$, one gets
\begin{equation}
    \frac{\Delta^2}{E_\alpha E_\beta (E_\alpha + E_\beta)} (\xi_\alpha-\xi_\beta)^2~g_{\alpha\beta}^{ij} =
    \Delta^2\bigg(\frac{1}{E_\beta}-\frac{1}{E_\alpha}\bigg)\frac{(\xi_\alpha-\xi_\beta)}{(\xi_\alpha+\xi_\beta)}~g_{\alpha\beta}^{ij} \,,
\end{equation}
where we are not writing the sum and the factor in front of it because they are irrelevant for the present discussion. In the above expression, the gap dependent quantities are only the pairing gap $\Delta$ and the quasi-particle energies $E_\alpha$,$E_\beta$. One can express the quasi-particle energies in terms of the single-particle ones and the gap, obtaining
\begin{equation} \label{eq:geometriccon}
    \Delta^2\bigg(\frac{1}{E_\beta}-\frac{1}{E_\alpha}\bigg)\frac{(\xi_\alpha-\xi_\beta)}{(\xi_\alpha+\xi_\beta)}~g_{\alpha\beta}^{ij} =
    \Delta \Bigg[\bigg(1+\frac{\xi_\beta^2}{\Delta^2}\bigg)^{-\frac{1}{2}}-~\bigg(1+\frac{\xi_\alpha^2}{\Delta^2}\bigg)^{-\frac{1}{2}}\Bigg]
    \frac{(\xi_\alpha-\xi_\beta)}{(\xi_\alpha+\xi_\beta)}~g_{\alpha\beta}^{ij} \,,    
\end{equation}
where the gap dependence has now been made explicit. At this point, one can study the limiting cases approximating the above expression. The integral over the Brillouin zone will sum over the contributions we will discuss, but since the pairing gap is constant, the gap dependence will survive the integration.
\begin{equation} \label{eq:appll}
    |\xi|\ll\Delta~~:~~\Delta\bigg(1+\frac{\xi^2}{\Delta^2}\bigg)^{-\frac{1}{2}} \simeq~~\Delta \,.
\end{equation}
Remembering that $\xi=\epsilon-\mu$, this limit contains the case of single-particle bands that cut the Fermi energy along a surface of the BZ, which are usually referred to as conduction bands. As it can be seen, one gets a linear dependence on the pairing gap, i.e. the contribution of each conduction band to the superfluid density (which is already finite because of the conventional contribution) increases linearly with $\Delta$.

\begin{figure}
    \centering
    \includegraphics[width=10cm]{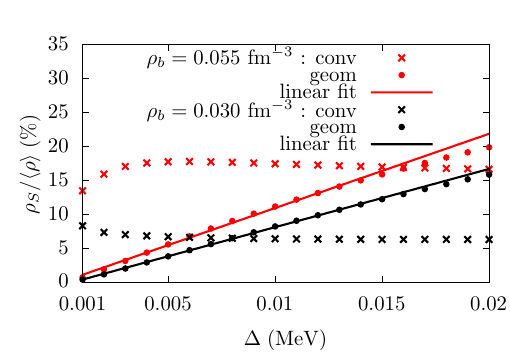}
    \caption{Conventional (crosses) and geometric (points) contributions to the superfluid fraction as functions of the pairing gap, computed respectively with Eqs.~(\ref{eq:supdens_conv}) and (\ref{eq:supdens_geom}) for two different baryon densities in the crystal phase of the inner crust of neutron stars (see \cite{Almirante25} for details). Lines are linear fits performed in the range $\Delta\in[0.001,0.015]$~MeV.}
    \label{fig:supdens}
\end{figure}

Since we are here interested in the case of small gaps, it is not necessary to go further in this expansion, and one can just look at the case in which single-particle energy and pairing gap are comparable. 
\begin{equation} \label{eq:appsimeq}
    |\xi|\simeq\Delta~~:~~\Delta\bigg(1+\frac{\xi^2}{\Delta^2}\bigg)^{-\frac{1}{2}} \simeq~~\frac{\Delta}{\sqrt{2}} \,.
\end{equation}
Also in the case in which single-particle energy and gap are comparable, a contribution linear in the gap emerges. It has to be noticed that in Eq.~(\ref{eq:geometriccon}) there is a minus sign between the contributions due to different bands, however the factors in front of the linear terms in Eq.~(\ref{eq:appll}) and Eq.~(\ref{eq:appsimeq}) are different, thus an exact cancellation does not occur. In this sense, one can expect that in systems in which many single-particle energies are close to the Fermi energy, the superfluid density will increase linearly when increasing the value of the gap. In Figure \ref{fig:supdens}, results from Eqs.~(\ref{eq:supdens_conv}) and (\ref{eq:supdens_geom}) are shown for the same configurations explored in \cite{Almirante25}, this is done solving the HF eigenvalue problem in Eq.~(\ref{eq:HFeigen}) and then computing the two contributions integrating over the BZ (see \cite{Almirante25} for details). The linear dependence of the geometric contribution on the magnitude of the gap is clearly visible. It has to be noticed that also in condensed matter systems a linear dependence of the superfluid density on the magnitude of the interaction strength has been found \cite{Julku16,Mojarro25}.    
\begin{equation}
    |\xi|\gg\Delta~~:~~\Delta\bigg(1+\frac{\xi^2}{\Delta^2}\bigg)^{-\frac{1}{2}} \simeq~~\frac{\Delta^2}{|\xi|} \,.
\end{equation}
Finally, the contribution due to bands with energy greater than the gap is quadratic in the latter. However, it has to be noticed that it is the gap times the parameter of the expansion, thus one can expect that it will be smaller than the contribution coming from the bands closer to the Fermi energy. Still, if there are no conduction bands (i.e. no bands cut the Fermi energy), this is the first contribution that arises from Eq.~(\ref{eq:supdens_geom}). If no single-particle bands cut the Fermi energy, the conduction density $\rho_c$ in the normal phase is zero. Moreover, the conventional contribution Eq.~(\ref{eq:supdens_conv}) is also negligible if there are no conduction bands. The interesting observation is that even in this case a non-zero superfluid density can still arise.

\section{Single-particle picture in linear response theory} \label{sec:singleparticle}
As explained in Section \ref{sec:formalism} our construction can be applied if the single-particle labels are good quantum numbers for the description of the system. Here we thus want to explore different ways of computing the correction to the density matrix $\rho'_{\alpha\beta}$.

\subsection{Linear response in HF}
In Section \ref{sec:emergence} we have mentioned the conduction density $\rho_c$ in the normal phase. In order to clarify what this means, and also to shed some light on the meaning of the two contributions to the superfluid density in Eq.~(\ref{eq:supdens}), let us compute the linear response to a perturbation $V'$ from the HF equations.

One can proceed in a way completely analogous to what has been done in the HFB case \cite{Almirante25}. Writing the perturbed eigenvalues and eigenvector in the form $\xi_\alpha+\xi_\alpha'$ and $|\alpha\rangle+|\alpha\rangle'$, respectively, the first-order corrections $\xi_\alpha'$ and $|\alpha\rangle'$ can be computed in perturbation theory \cite{Sakurai} as
\begin{equation} \label{eq:perturbedHF}
    \xi_\alpha'=V'_{\alpha\alpha}
    ~~~,~~~
    |\alpha\rangle'=\sum_{\beta\neq\alpha}|\beta\rangle\frac{V'_{\beta\alpha}}{\xi_\alpha-\xi_\beta} \,.
\end{equation}
The unperturbed density matrix is given by
\begin{equation}
    \rho=\sum_\alpha \Theta(-\xi_\alpha)|\alpha\rangle\langle\alpha| \,,
\end{equation}
and writing the perturbed one as $\rho+\rho'$, one finds for the linear perturbative correction
\begin{equation} \label{eq:densitymatrix_perturbation}
    \rho'=\sum_\alpha \Big[-\delta(\xi_\alpha)\xi_\alpha'|\alpha\rangle\langle\alpha|+
    \Theta(-\xi_\alpha)\big(|\alpha\rangle'\langle\alpha|+|\alpha\rangle\langle\alpha|'\big)\Big] \,,
\end{equation}
where the $\delta$ function comes from the correction to the single-particle energies, while the second term comes from the eigenstate corrections. Notice that the $\delta$ function does not appear in the correction to the density matrix computed through the HFB equations \cite{Almirante25,Migdal59}. This is due to the fact that in the HFB case there is a gap in the quasi-particle spectrum.  

Thus, for the elements of the density matrix in the unperturbed basis, one gets
\begin{equation}
    \rho'_{\alpha\beta}=\langle\alpha|\rho'|\beta\rangle=-\delta(\xi_\alpha)V'_{\alpha\alpha}~\delta_{\alpha\beta} +
    \frac{\Theta(-\xi_\alpha)-\Theta(-\xi_\beta)}{\xi_\alpha-\xi_\beta}V'_{\alpha\beta}(1-\delta_{\alpha\beta}) \,.
\end{equation}
Now, one can write explicitly the perturbation $V'=-\vv\cdot\pv$, and rearranging the average momentum density Eq.~(\ref{eq:current}) as 
\begin{equation}\label{eq:responsej_HF}
    \langle j^i\rangle = \sum_{ij} R^{ij}~v^j\,,
\end{equation}
one gets for the $R$ tensor
\begin{equation}\label{eq:responseR-HF}
    R^{ij} =-2\!\int_{\text{BZ}}\frac{d^3k}{(2\pi)^3}
    \Bigg[\sum_{\alpha} 
    (-\delta(\xi_\alpha))
    p^i_{\alpha\alpha} p^j_{\alpha\alpha}
    +
    \sum_{\alpha\neq\beta} 
    \frac{\Theta(-\xi_\alpha)-\Theta(-\xi_\beta)}{\xi_\alpha-\xi_\beta}
    p^i_{\alpha\beta} p^j_{\beta\alpha}\Bigg] \,.
\end{equation}
Noticing that $-\delta(\xi_\alpha)=\partial\Theta(-\xi_\alpha)/\partial\xi_\alpha$, one obtains the same expression for the $R$ tensor as in \cite{Almirante25} but with the HF occupation numbers $\Theta(-\xi_\alpha)$, instead of the BCS occupation numbers $v_\alpha^2$. In this respect, one can follow the same steps as in \cite{Almirante25}, and gets
\begin{equation}\label{eq:responseR-HF-final}
  R^{ij} = 2 m\int_{\text{BZ}}\frac{d^3k}{(2\pi)^3}\sum_{\alpha} \Theta(-\xi_\alpha)\, \delta^{ij} = m\langle\rho\rangle\,\delta^{ij} \,.
\end{equation}
It is perfectly natural that the HF response, computed for a stationary flow as it has been done here, gives the total density. This is because a response computed time-independently can only describe an equilibrium picture, thus in the HF case the collective rigid motion of all the particles.

However, already at this level, one can recognize two contributions to the response, one coming from the perturbed single-particle energies, and the other from the perturbed eigenstates. If the system was homogeneous, only the first term in Eq.~(\ref{eq:densitymatrix_perturbation}) would contribute, this is what is usually referred as the conduction density $\rho_c$. If instead the system is not homogeneous, then the second term starts to increase and the first to decrease. In the case in which no single-particle energies cut the Fermi energy, i.e. all bands are completely filled, the conduction density $\rho_c$ is zero, and the total density comes from the second term only.

In this sense, one can interpret the conduction density $\rho_c$ as the particles that are actually ``free'' to move, i.e., in electronic systems in the normal phase, the electrons that can be put in motion by applying an electric field. Concerning the second term in Eq.~(\ref{eq:responseR-HF}), it describes the particles that in our response rigidly follow all the others, but in a normal electronic system would not participate in conduction.

From this point of view, one can get an intuitive understanding of the geometric contribution to the superfluid density discussed above. In the case in which the conduction density $\rho_c$ is negligible, if the geometric contribution is different from zero, a collective motion of particles can still be sustained. This can happen under the condition that pairing correlations are strong enough that they can mix otherwise empty and filled bands, i.e. the ones that contribute to the second term of Eq.~(\ref{eq:responseR-HF}).  

\subsection{Linear response in HFB with perturbed HF single-particle states}
The idea here is to perform perturbation theory on the HFB equations, but accounting only for the correction to the single-particle states. This means that the density matrix will be computed through the generalized density matrix, while the correction to the quasi-particle states will be approximated only through their relation with the single-particle states. 

The unperturbed HFB matrix has eigenvalues $E_{\alpha}$ and $-E_{\alpha}$. We denote the corresponding unperturbed eigenvectors with positive and negative energies by $|\alpha_+\rangle$ and $|\alpha_-\rangle$, respectively, and they are given by
\begin{equation}\label{eq:defA}
    |\alpha_+\rangle = 
    \begin{pmatrix}u_{\alpha}\\-v_{\alpha}\end{pmatrix}
    |\alpha\rangle\,,\quad
    |\alpha_-\rangle = 
    \begin{pmatrix} v_{\alpha}\\u_{\alpha}\end{pmatrix}
    |\alpha\rangle\,.
\end{equation}
We write the (negative-energy) perturbed eigenvector in the form $|\alpha_-\rangle+|\alpha_-\rangle'$, but instead of using the complete set of HFB eigenstates, we use only the perturbed HF single-particle states Eq.~(\ref{eq:perturbedHF}), namely
\begin{equation}\label{eq:correctionA}
    |\alpha_-\rangle' = 
    \begin{pmatrix} v_{\alpha}\\u_{\alpha}\end{pmatrix}
    \sum_{\beta\neq\alpha}|\beta\rangle\frac{V'_{\beta\alpha}}{\xi_\alpha-\xi_\beta} \,.
\end{equation}
The unperturbed generalized density matrix is given by
\begin{equation}
	\mathcal{R}=\sum_{\alpha} |\alpha_-\rangle\langle\alpha_-| \,,
\end{equation}
and writing the perturbed one as $\mathcal{R}+\mathcal{R}'$, one finds for the linear pertubative correction
\begin{equation}
    \mathcal{R}' = \sum_{\alpha} (|\alpha_-\rangle'\langle\alpha_-|
    +|\alpha_-\rangle\langle\alpha_-|')\,.
\end{equation}
Using Eqs.~\eqref{eq:defA} and \eqref{eq:correctionA} and the hermiticity of $V'$ (i.e., $V'_{\alpha\beta}=V'_{\beta\alpha}$), one finds
\begin{equation}
    \rho'_{\alpha\beta} =
    \langle\alpha|\mathcal{R}_{11}|\beta\rangle\\
    =\frac{v_\alpha^2-v_\beta^2}{\xi_\alpha-\xi_\beta}V'_{\alpha\beta}(1-\delta_{\alpha\beta}) \,, 
\end{equation}
where $\mathcal{R}_{11}$ is the upper left element of the generalized density matrix \cite{Almirante24,Almirante25}, which is a $2\times 2$ matrix due to the vector nature of the HFB eigenvectors Eq.~(\ref{eq:defA}).

To go on with the calculation, the density matrix can be rewritten as
\begin{equation}
    \rho'_{\alpha\beta} =
    \Big[\frac{\partial v_\alpha^2}{\partial \xi_\alpha} V'_{\alpha\alpha}\delta_{\alpha\beta}+
    \frac{v_\alpha^2-v_\beta^2}{\xi_\alpha-\xi_\beta}V'_{\alpha\beta}(1-\delta_{\alpha\beta})\Big] -
    \frac{\partial v_\alpha^2}{\partial \xi_\alpha} V'_{\alpha\alpha}\delta_{\alpha\beta} \,, 
\end{equation}
where we summed and subtracted the same quantity $(\partial v_\alpha^2/\partial \xi_\alpha) V'_{\alpha\alpha}\delta_{\alpha\beta}$. 

Now we can rewrite explicitly the average momentum density
\begin{equation}\label{eq:responseRS}
    \langle j^i\rangle = \sum_{ij} \big(R^{ij} - S^{ij}\big) v^j\,,
\end{equation}
where $R$ and $S$ are tensors given by
\begin{equation}\label{eq:responseR-rewritten}
    R^{ij} =-2\!\int_{\text{BZ}}\frac{d^3k}{(2\pi)^3}
    \Bigg[\sum_{\alpha} 
    \frac{\partial v^2_{\alpha}}{\partial \xi_{\alpha}}
    p^i_{\alpha\alpha} p^j_{\alpha\alpha}
    +
    \sum_{\alpha\neq\beta} 
    \frac{v^2_{\alpha}-v^2_{\beta}}{\xi_{\alpha}-\xi_{\beta}}
    p^i_{\alpha\beta} p^j_{\beta\alpha}\Bigg] \,,
\end{equation}
\begin{equation}
    S^{ij} =\int_{\text{BZ}}\frac{d^3k}{(2\pi)^3}\sum_{\alpha}
    \frac{\Delta^2}{E_{\alpha}^3}
    p_{\alpha\alpha}^i p_{\alpha\alpha}^j \,.
\end{equation}
The $R$ tensor is what has been found in \cite{Almirante25}, and it can be shown that
\begin{equation}\label{eq:responseR-final}
  R^{ij} = 2 m\int_{\text{BZ}}\frac{d^3k}{(2\pi)^3}\sum_{\alpha} v_{\alpha\kv}^2\, \delta^{ij} = m\langle\rho\rangle\,\delta^{ij} \,.
\end{equation}
The $S$ tensor has instead been rewritten noticing that
\begin{equation}
    \frac{\partial v^2_{\alpha}}{\partial \xi_{\alpha}}=-\frac{\Delta^2}{2E_\alpha^3} \,.
\end{equation}
Comparing the resulting average momentum density with Eq.~(\ref{eq:currenttwofluid}), and using Eq.~(\ref{eq:momentum}), one gets
\begin{equation}
    \rho_S^{ij} =m\int_{\text{BZ}}\frac{d^3k}{(2\pi)^3}\sum_{\alpha}
    \frac{\Delta^2}{E_{\alpha}^3}
    \frac{\partial\xi_{\alpha}}{\partial k^i} \frac{\partial\xi_{\alpha}}{\partial k^j} \,.    
\end{equation}
As it can be seen the resulting superfluid density comes now from the sole conventional contribution.

The important point here is that the geometric contribution does not appear because the perturbation has been applied only to the single-particle states, leaving the Bogoliubov coefficients $u$ and $v$ unperturbed in the expression for the correction to the quasi-particle states in Eq.~(\ref{eq:correctionA}). In this sense the intuitive picture that one gets from the HF response seems to be meaningful: the geometric contribution comes from a mixing between empty and filled bands. By neglecting the correction to the Bogoliubov coefficients, one can get only the conventional term. 

This result also clarifies within a comprehensive framework what was missing in previous calculations in linear response theory for the superfluid density \cite{Carter05A,Chamel24}. As noticed in \cite{Almirante25} the problem was not the BCS approximation, but the neglect of the non-diagonal terms in the average momentum density. Here we showed that this neglect follows from a derivation where one does not take into account the mixing between Bogoliubov coefficients of different bands, which comes directly from the perturbation. 

\section{Conclusions} \label{sec:conclusions}
Within the linear response theory framework, we analyzed the response of a 3D periodic system to a stationary flow. The study is performed assuming that the system can be described within a mean-field approach, specifically requiring that the single-particle labels are good quantum numbers. In the HF theory this is a trivial statement, while in the HFB case to satisfy this requirement one has to start from the BCS approximation.

In the presence of superfluidity, this approach allows one to compute the superfluid density of the system. In the HFB case, this was already done in \cite{Almirante25}, and here we showed how the geometric contribution emerges in the case in which the system is characterized by a highly non-trivial band structure, i.e. when many bands cut the Fermi energy. This case is of particular interest for the physics of the inner crust of neutron stars, where matter is expected to be much more dense than in the condensed matter systems we can study in laboratories. In particular, it has been found that a linear dependence on the magnitude of the pairing gap is expected, in agreement with the numerical evaluation of Eq.~(\ref{eq:supdens}) performed in \cite{Almirante25} and with the full HFB calculations for the 2D and 3D periodic cases of \cite{Almirante24A,Almirante25A}. In order to compare the results for the superfluid density at constant gap with the full HFB calculations, one has to average the HFB gaps, as done in \cite{Almirante25,Almirante25A}.

Moreover, we showed that also in the HF case the response to a stationary flow presents two contributions. The first coming from each band that cuts the Fermy energy, the so-called conduction bands, which is usually interpreted as the conduction density. The second coming from a sum over different bands, which is needed to get the total density, since in HF all the particle have to rigidly move as a consequence of the stationary flow.

Then, we find that in our framework the neglect of the perturbative correction to the occupation numbers leads to the sole conventional contribution to the superfluid density. This shed some light on the origin of the geometric contribution in many bands systems: if the pairing correlations are big enough a mixing between empty and filled bands allows to sustain the collective motion of particles.

However, it has to be noticed that this mixing is allowed only if the momentum operator connects the interested single-particle states, as it is clear by the presence of the off-diagonal momentum matrix elements in Eq.~(\ref{eq:supdens_geom}). The physical meaning of these processes has been started to be discussed in the context of electronic systems \cite{Hu25}.

In this work we started to clarify how the geometric contribution to the superfluid density is involved in the physics of the inner crust of neutron stars. As it was shown in \cite{Martin16}, with superfluid fractions as we found with the inclusion of the geometric contribution, the glitches of the Vela pulsar can be explained with the superfluidity of the crust alone, and it is not necessary to modify the glitch models to include superfluidity in the core as suggested in \cite{Andersson12}. Having a reliable estimate of the superfluid density, one can study implications for the low-lying phonons \cite{Pethick10,Chamel13B,Durel18} and hence for the thermal evolution of neutron stars \cite{Page12}, as well as for the frequencies of star oscillation modes \cite{Andersson02,Sotani12,Tews17}.

However, it is important to notice that here our picture is purely stationary. On one hand, this is what allows us to extract the superfluid density. On the other hand, in such an approach the dynamical features of the system remain unknown. In this sense, this work is only the starting point for the exploration of the importance that the geometric contribution could have in the inner crust of neutron stars.

Apart from the astrophysical importance of understanding how superfluidity affects the properties of the inner crust of neutron stars, it is of great interest to know in what conditions one can compare it to the condensed matter systems that we can study experimentally on Earth. These kind of comparisons, while difficult because of the different nature of these systems, are of fundamental importance since they could help to explore astrophysical processes that otherwise would be unreachable.

\acknowledgments{The author thanks Michael Urban for fundamental discussions needed to accomplish this work and critical reading of the manuscript.}

\bibliography{refs}
\end{document}